\documentclass{article}

\usepackage[final]{nips_2017}

\usepackage[utf8]{inputenc} 
\usepackage[T1]{fontenc}    
\usepackage[hidelinks]{hyperref}       
\usepackage{url}            
\usepackage{booktabs}       
\usepackage{amsfonts}       
\usepackage{nicefrac}       
\usepackage{microtype}      
\usepackage{graphicx}
\graphicspath{{figures/}}
\usepackage{lipsum}
\usepackage{newfloat}
\DeclareFloatingEnvironment[name={Supplementary Figure}]{suppfigure}
\usepackage{amsmath}
\usepackage{textcomp}

\title{Exploiting oddsmaker bias to improve the prediction of NFL outcomes}

\author{
  Erik J. Schlicht\thanks{Website: \url{http:/schlicht.org}} \\
  Computational Cognition Group, LLC\\
  St. Paul, MN 55123 \\
  \texttt{ejs@c2-g.com} \\
}

\begin{document}

\maketitle

\begin{abstract}
Accurately predicting the outcome of sporting events has been a goal for many groups who seek to maximize profit. What makes this challenging is that the outcome of an event can be influenced by many factors that dynamically change across time.  Oddsmakers attempt to estimate these factors by using both algorithmic and subjective methods to set the spread. However, it is well-known that both human and algorithmic decision-making can be biased, so this paper explores if oddsmaker biases can be used in an exploitative manner, in order to improve the prediction of NFL game outcomes.  Real-world gambling data was used to train and test different predictive models under varying assumptions. The results show that methods that leverage oddsmaker biases in an exploitative manner perform best under the conditions tested in this paper.  These findings suggest that leveraging human and algorithmic decision biases in an exploitative manner may be useful for predicting the outcomes of competitive events,  and could lead to increased profit for those who have financial interest in the outcomes. \\

Key Words: Bias Exploitation, Forecasting, Shannon Entropy, Sports Gambling, NFL, Risk \\


\end{abstract}

\section{Introduction}
\label{intro}
Sporting events are widely popular and of interest to many people across several cultures.  Although most attend sporting events for entertainment purposes, some have financial investment in the outcome of the game.  These groups have a keen interest in accurately predicting the results of such events, so it is not surprising that several attempts have been made by researchers to develop methods and algorithms that predict the outcome of various sporting events (\cite{Aoki:2017,Dani:06,Kampakis:2014,Kampakis:2015,Kiraly:2017,Knowlton:2017,Pelechrinis:2017}).

Oddsmakers are one group who attempt to accurately predict the result of games in order to maximize their profit.  They accomplish this by setting the \emph{spread} at a value that encourages equal betting by gamblers for each team in the competition\footnote{Description of spread estimation process: \url{https://tinyurl.com/p5yxekh}}. What makes this challenging is that the outcome of an event can be influenced by many factors that dynamically change across time (\cite{Aoki:2017}), so the spread can be considered a high-level metric that captures these variables.

\begin{figure}[ht]
\vskip 0.2in
\begin{center}
\includegraphics[width=.8\columnwidth]{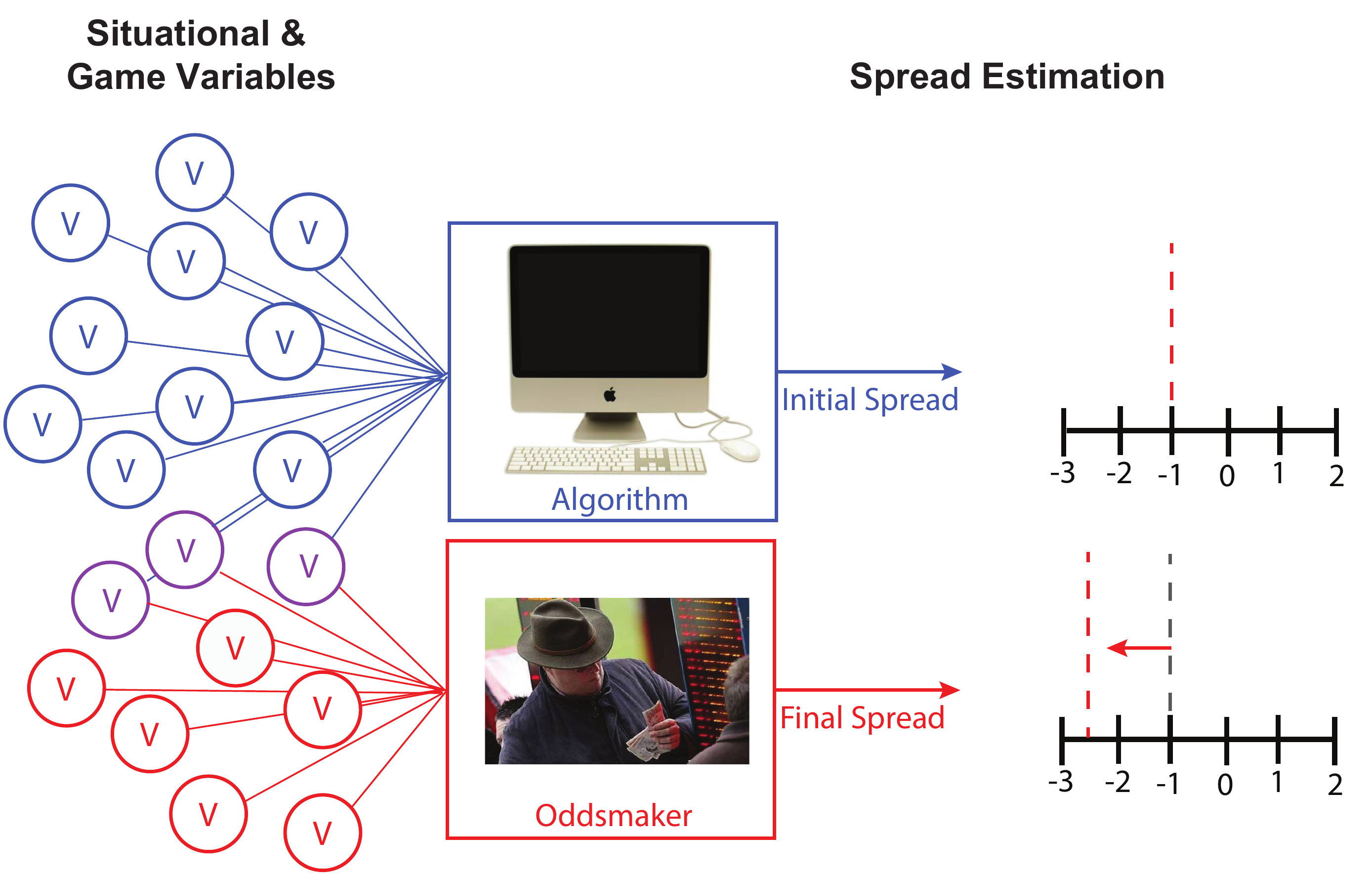}
\caption{Diagram depicting the process by which game spreads are estimated}
\label{fig:fig1}
\end{center}
\vskip -0.2in
\end{figure}

Due to the difficulty of setting the spread, oddmakers will often leverage algorithms to set the initial value and manually adjust it as more information becomes available (Figure \ref{fig:fig1}).  This process introduces an interesting issue, since it is well-known that both algorithmic performance (\cite{Datta:15,Hajian:16,Sweeney:13}) and human decision-making (\cite{Camerer:02,David:11,Gilovich:93,Gilovich:02,Haghighat:13,Kahneman:79,Kahneman:82,Kahneman:12,Schlicht:10}) can be biased.  Therefore, oddsmakers may be (implicitly) providing reliable information about the outcome of a sporting event to gamblers through the spread.  Using poker terminology, the spread may provide a \emph{tell} that gamblers can exploit to improve predictive performance.

This study is the first to investigate if oddsmaker bias can be exploited to improve the prediction of outcomes in NFL games. Previous research exclusively leveraged game and situational features (e.g., historic win percent, margin of victory) in their predictive models (\cite{Aoki:2017,David:11,Haghighat:13,Kiraly:2017,Knowlton:2017,Pelechrinis:2017}), or aggregated expert predictions to improve model performance (\cite{Dani:06}). This effort will extend existing research by systematically exploring if spread biases provide reliable information about game outcomes.

First, Section \ref{timodel} explores how models that exploit decision biases perform against models that do not leverage this information, under conditions of temporally-independent data.  Then, Section \ref{tdmodel} will investigate how the same models perform under temporally-dependent conditions, that better reflect the nature of real-world gambling. Before performance is considered, the next section will describe how the data used in this effort was leveraged to train and evaluate model performance under the conditions considered in this paper.

\section{Methods and Results}
\label{methods}
In order to investigate if oddsmaker bias can be exploited to improve the prediction of NFL outcomes, real-world gambling data need to be acquired to train and evaluate the predictive models.  Section \ref{data} describes how the data used in this effort was obtained and filtered.

Section \ref{timodel} will outline how the data was leveraged to train and test predictive models under conditions in which the date of the game is ignored (i.e., temporally-independent data conditions). Then, Section \ref{tdmodel} outlines how the models perform under situations where the data used to train and test each model are sensitive to dates in which the games were played (i.e., temporally-dependent conditions).

\subsection{Real-World Data}
\label{data}
\begin{figure}[ht]
\vskip 0.2in
\begin{center}
\includegraphics[width=\columnwidth]{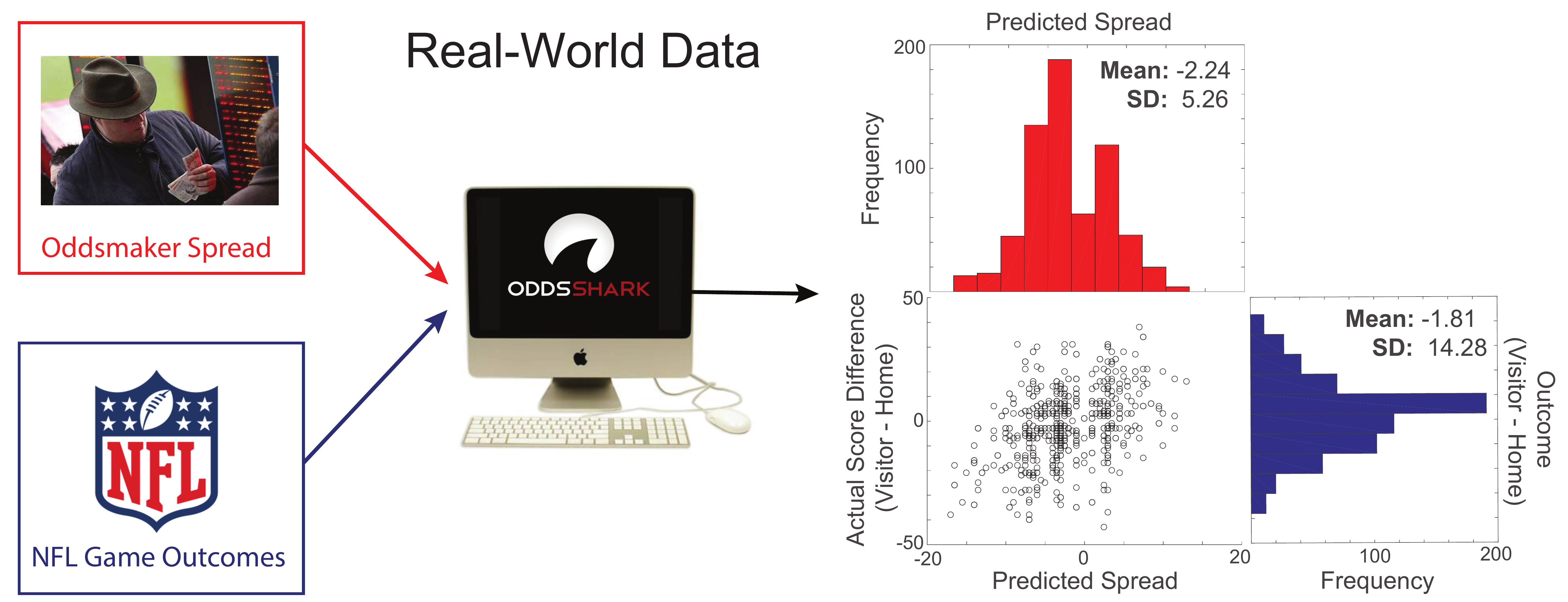}
\caption{Both spread and NFL outcome data were obtained from an online gaming site}
\label{fig:fig2}
\end{center}
\vskip -0.2in
\end{figure}

To explore if spread bias can be exploited to improve the prediction of NFL outcomes, real-world gaming data are required that contain both spread and outcome information (Figure \ref{fig:fig2}).  This study leveraged data obtained from an online gambling site\footnote{OddsShark NFL Database (uploaded on 10.11.17): \url{http://www.oddsshark.com/nfl/database}}.

The maximum amount of data that could be obtained from the database had an upper-bound of 30 samples per team.  Therefore, the maximum number of games were uploaded, which resulted in 960 unfiltered samples (30 samples x 32 NFL teams). The samples obtained were all regular-season conference games that took place between December, 2014 and October, 2017. Since the database produced duplicates for some games, they were removed which resulted in 648 unique samples that contained both NFL game outcome (Visitor Score - Home Score) and the final spread.  The next section describes how the data were used to train the predictive models used in this study.

\subsection{Temporally-Independent (TI) Models}
\label{timodel}

The goal of this study is to systematically explore if spread estimates provide reliable information about game outcomes. To accomplish this goal, it is necessary that the probability of an outcome ($o$) given a valid spread ($s$) be estimated ($p(o \mid s)$).

For the temporally-independent (TI) models described first, this density is estimated through a training procedure that does not consider the date in which games were played.  In this respect, models can be trained using data from more recent games, and evaluated on games that were played further in the past.  Obviously, this doesn't directly approximate real-world context, but it allows for our analysis to simulate hundreds of games to provide insight into the variability of model performance.  The next section describes how models were trained and evaluated under TI conditions.

\subsubsection{TI-Model Training}
\label{titrain}

\begin{figure}[ht]
\vskip 0.2in
\begin{center}
\includegraphics[width=\columnwidth]{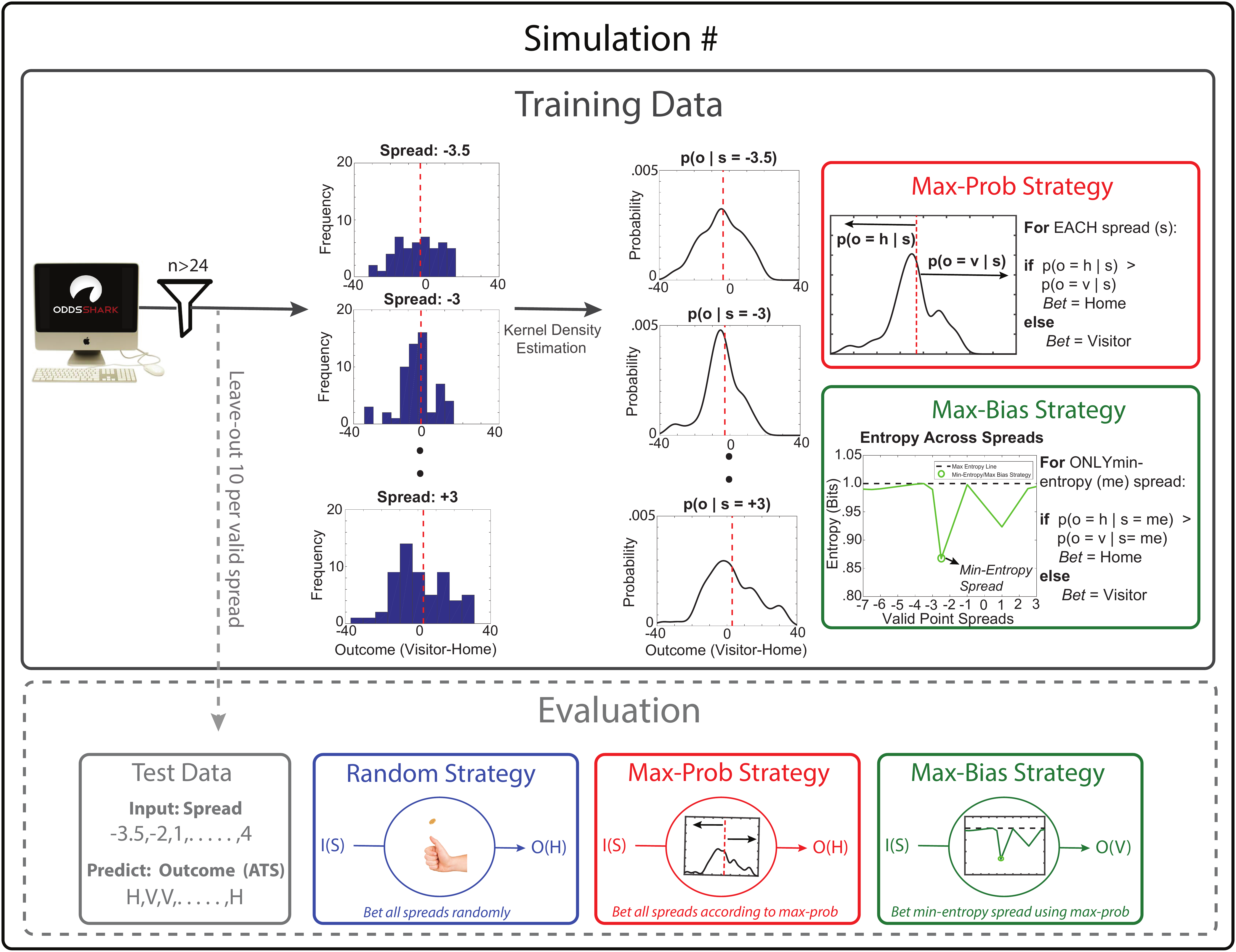}
\caption{Process used to train and evaluate temporally-independent methods}
\label{fig:fig3}
\end{center}
\vskip -0.2in
\end{figure}

To explore if oddsmaker bias can be exploited to improve the prediction of NFL outcomes, candidate models need to be defined in order to compare their relative performance across the TI conditions considered in this section.  The most straightforward model that defines a lower-bound on performance is the \emph{Random-Guess} model.  Since there are two possible outcomes (i.e., home wins or visitor wins), the Random-Guess model decision ($D_{rg}$) is analogous to a coin-flip (ignoring ties which are uncommon in the NFL).  Notice that the Random-Guess model does not use potential information about the relationship between spread and outcomes when making its prediction:

\begin{equation}
\label{eq:eq1}
  D_{rg}=\left\{
  \begin{array}{@{}ll@{}}
    \text{Visitor}, & \text{if}\ rand < \frac{1}{N_o} \\
    \text{Home}, & \text{otherwise}
  \end{array}\right.
\end{equation}

where $rand$ is a random number generation sample (0-1) and $N_o$ is the number of possible outcomes ($N_o=2$).

Another model that is considered in this paper is termed the \emph{Maximum-Probability} model.  The Max-Probability model\textquotesingle s predictions ($D_{mp}^{s}$) select the team that maximizes the probability of an after-the-spread (ATS) win for either the home ($h$) or visiting ($v$) teams. Predictions are made for each valid spread ($s$):

\begin{equation}
\label{eq:eq2}
  D_{mp}^{s}=\left\{
  \begin{array}{@{}ll@{}}
    \text{Visitor}, & \text{if}\ p(o = v \mid s) > p(o = h \mid s) \\
    \text{Home}, & \text{otherwise}
  \end{array}\right.
\end{equation}

The validity of a spread corresponds to the number of samples available at each spread value. Under temporally-independent (TI) conditions considered in this section, valid spreads were required to have at least 25 outcome samples (Figure \ref{fig:fig3}). The data filtering process resulted in 7 valid spread values ($s$), for which histograms are shown in Supplementary Figure \ref{fig:figS1}.

This filtering was performed since the data need to be separated into training data used to estimate $p(o \mid s)$), and test data that was used to evaluate predictive performance. Figure \ref{fig:fig3} shows the process by which the data were separated.  For each of the N-simulations (N=200), 10 data samples per valid spread were held-out as test data, resulting in 14,000 test samples (10 samples x 7 valid spreads x 200 simulations), or 2000 test samples per valid spread.

Since we required at least 25 samples per spread, this assures that there is a minimum of 15 training samples that can be used to estimate $p(o \mid s)$, per simulation. Supplementary Figure \ref{fig:figS2} shows the estimated pdf for each of the 7 valid spead values.  Notice that the date of the game was not considered during the separation of data between train and test conditions, which is why it is termed temporally-independent (TI) conditions.

In order to estimate the densities needed to make maximum-probability predictions ($ D_{mp}^{s}$), we used kernel density estimation (with width = 4) to estimate the quantities of interest (Figure \ref{fig:fig3}):

\begin{equation}
\begin{aligned}
    p(o = h \mid s) &= \int_{-\inf}^{s} p(o \mid s)ds
    &= \sum_{\min s_a}^{s} p(o \mid s_a)
\end{aligned}
\end{equation}

where $s$ is the valid spread value, $s_a$ are the quantized outcome values used during the kernel density estimation (-40 to +40), and  $p(o = v \mid s) = 1 - p(o = h \mid s)$.

Notice that both Random-Guess and Maximum-Probability models make predictions across each of the valid spread values, and do not take into account oddsmaker bias.  However, the main objective of this paper is to explore how decision-bias can be exploited to improve NFL outcome prediction, so models that accomplish this objective are considered next.

A first step to exploiting decision-bias is to find a method that is capable of quantitatively identifying biases. Since bias can be considered the amount of information the spread provides about the outcome, we can use Shannon entropy (\cite{Shannon:48}) as an effective metric to for this purpose. In this respect, Shannon entropy ($H(s) = - \sum_t p(o = t|s) \log_2(p(o = t|s))$) reflects the amount of uncertainty associated with information carried by spread estimates in their ability to predict outcomes (for each team ($t$)).  Stated another way, decisions that are strongly biased will correspond to minimum-entropy spreads.  Spreads that exceed some threshold-level of entropy (.95 in our work) are then identified and considered biased.  These biased spreads can then exploited to improve predictive performance.

The \emph{Min-Entropy} model essentially makes the same predictions ($D_{me}^{s^*_1}$) as the Max-Probability model for a given spread.  However, whereas the Max-Probability model wagers across all valid spreads, the Min-Entropy model only wagers for spreads that are maximally biased (i.e., minimum entropy spread ($s^*_1$)).  In this respect, the Min-Entropy approach determines \emph{which spread} to select through Shannon entropy, and then uses a Maximum-Probability approach to decide \emph{which team} to choose (Figure \ref{fig:fig3}):

\begin{equation}
\label{eq:eq3}
  D_{me}^{s^*_1}=\left\{
  \begin{array}{@{}ll@{}}
    \text{Visitor}, & \text{if}\ p(o = v \mid s^*_1) > p(o = h \mid s^*_1) \\
    \text{Home}, & \text{otherwise}
  \end{array}\right.
\end{equation}

A more general version of the Min-Entropy model is the \emph{k-Lowest Entropy} model.  Whereas, the Min-Entropy model wagers only on the spread with the largest bias, the k-Lowest Entropy model wagers ($D_{ke}^{s^*_{1:k}}$) on all (k) spreads that are below the threshold-level of entropy (.95 in this paper). Once the k-lowest entropy spreads ($s^*_{1:k}$) are identified, those spreads are selected and teams are chosen for each spread based on Max-Probability methods.

\begin{equation}
\label{eq:eq4}
  D_{ke}^{s^*_{1:k}}=\left\{
  \begin{array}{@{}ll@{}}
    \text{Visitor}, & \text{if}\ p(o = v \mid s^*_{1:k}) > p(o = h \mid s^*_{1:k}) \\
    \text{Home}, & \text{otherwise}
  \end{array}\right.
\end{equation}

This section outlined how each of the candidate models (Random-Guess, Max-Probability, Min-Entropy, and k-Lowest Entropy) were trained and evaluated under TI conditions.  The next section discusses the results obtained through 200 simulations that were conducted under these TI conditions.

\subsubsection{TI-Model Evaluation}
\label{titest}
\begin{figure}[ht]
\vskip 0.2in
\begin{center}
\includegraphics[width=\columnwidth]{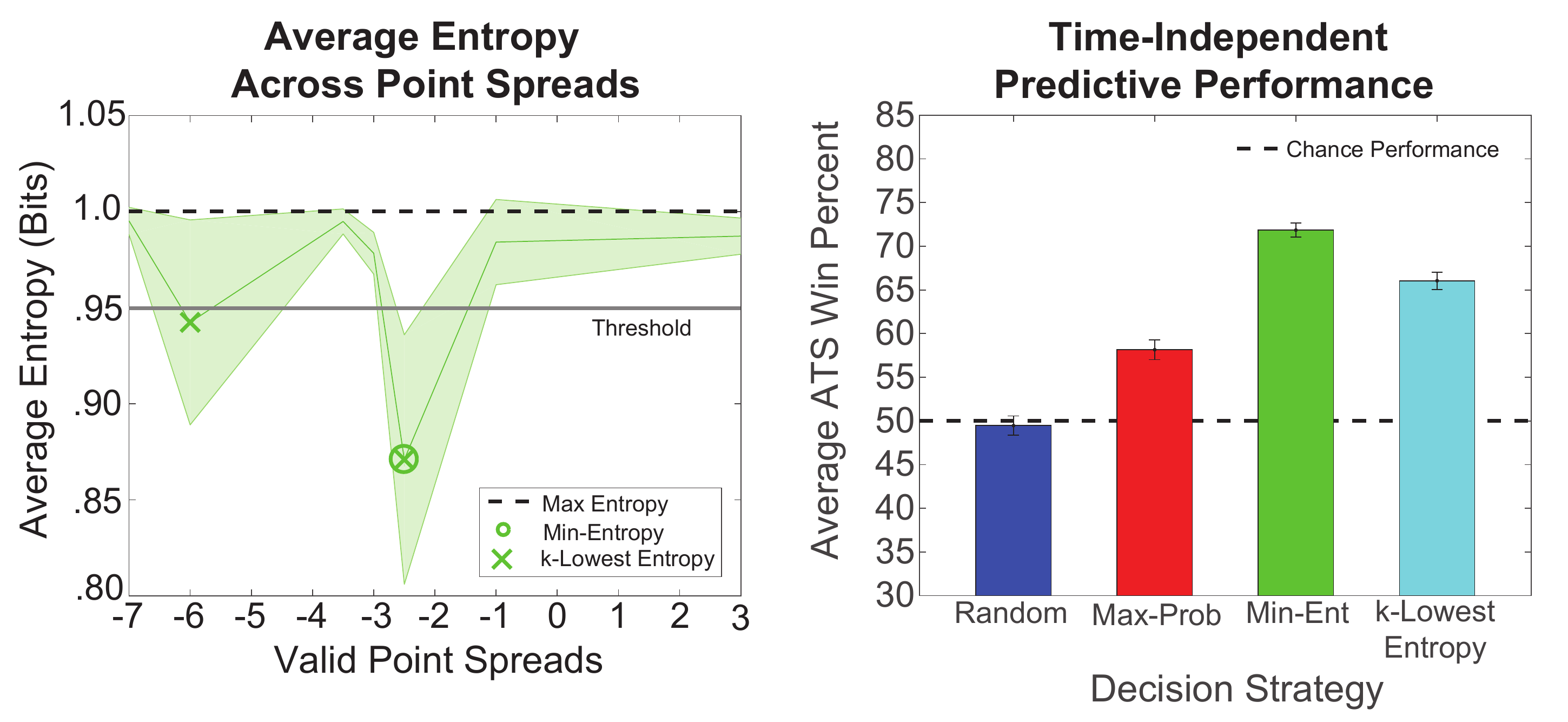}
\caption{Diagram depicting the average entropy ($\pm$ 1 SD) associated with each spread \emph{(Left)} and the mean predictive performance of each strategy ($\pm$ 1 SEM) on test data under temporally-independent conditions \emph{(Right)}.}
\label{fig:fig4}
\end{center}
\vskip -0.2in
\end{figure}

Using the procedure outlined in Section \ref{titrain}, model performance was evaluated with the test data held-out during each training iteration.  As mentioned above, this resulted in 2000 test samples for each of the 7 valid spread values.

The Random-Guess model effectively flipped a coin for each of the 2,000 test samples across each of the 7 valid spreads.  The actual NFL game outcomes from the test data was compared to Random-Guess model predictions, and resulted in average performance (Mean: 49.46\%; SEM: 1.11) that was close to expected levels of chance (Figure \ref{fig:fig4}, Right).

Predictions from the Max-Probability model were compared to actual NFL outcomes across all valid spreads for the test data. The results show a slight improvement in average predictive performance (Mean: 58.15\%; SEM: 1.13), above chance performance (Figure \ref{fig:fig4}, Right).

Finally, performance of the models that exploit oddsmaker bias were examined by comparing predictions from the Min-Entropy model to the 2,000 actual NFL outcomes for \emph{only the test data of the minimum entropy spread} (Figure \ref{fig:fig4}, Left).  Similarly, the k-Lowest Entropy model predictions were compared to each of the (k=2) valid spreads that fall below the entropy threshold (.95), resulting in 4,000 (2 spreads x 2,000 samples) comparisons.  As Figure \ref{fig:fig4} (Right) shows, both Min-Entropy (Mean: 71.85\%; SEM: .82) and k-Lowest Entropy (Mean: 66\%; SEM:.99) outperform the other two approaches that do not leverage biases in oddsmaker decisions.

Although the results under these temporally-independent conditions appear promising, it would be remiss not to evaluate how the entropy-based models perform under conditions that better reflect real-world gambling conditions. The next section will provide such an evaluation under temporally-dependent conditions that better reflect these natural conditions.

\subsection{Temporally-Dependent (TD) Models}
\label{tdmodel}

The previous section was able to simulate hundreds of trials in order to evaluate the performance of different models under TI conditions.  This section will strike the opposite balance by using historic data in an attempt to predict more recent outcomes.  Although these temporally-dependent (TD) conditions accurately reflect the temporal components of real-life gambling, they only allow for performance to be evaluated one time.

The hope is by leveraging these complimentary approaches for model evaluation, insight can be gained into the effectiveness expected from each of the candidate approaches if utilized outside of this study.

\subsubsection{TD-Model Training}
\label{tdtrain}

The candidate TD models are identical to those outlined in Section \ref{titrain}, and the training procedure is also similar to that reflected in Figure \ref{fig:fig3}, with a couple key differences.

First, instead of simulating over multiple iterations requiring train and test data to be randomly selected for each iteration, TD conditions separate the data only once. This one-time separation is based exclusively on the date in which the games were played.  More specifically, the test data exclusively included data from games played in the year 2017 (n=85), whereas training data included games from years prior to 2017 (2014-2016; n = 648-85 = 563). Implicitly, this reflects a situation where the gambler trained the model between the 2016 and 2017 seasons, and did not update the model as results became available.  Therefore, this can be considered a conservative estimate on model performance under TD conditions.

A second difference was in the definition of what constitutes a valid spread. The minimum number of samples required per spread was reduced to 15 (instead of 25).  Remember, this threshold determined the amount of training data used to estimate $p(o \mid s)$, but unlike TI conditions, no additional data in the TD condition need to be removed for testing. Effectively, this filtering reduction balanced the minimum training samples required for a valid spread between TI and TD conditions.

This filtering process resulted in 12 valid spreads which produced a total of 54 (of the possible 85) test samples across each of the valid spreads. Therefore, there is relatively limited data to evaluate performance due to the TD constraints, and Section \ref{tdtest} will overview the performance of each model under the conditions described in this section.

\subsubsection{TD-Model Evaluation}
\label{tdtest}
\begin{figure}[ht]
\vskip 0.2in
\begin{center}
\includegraphics[width=\columnwidth]{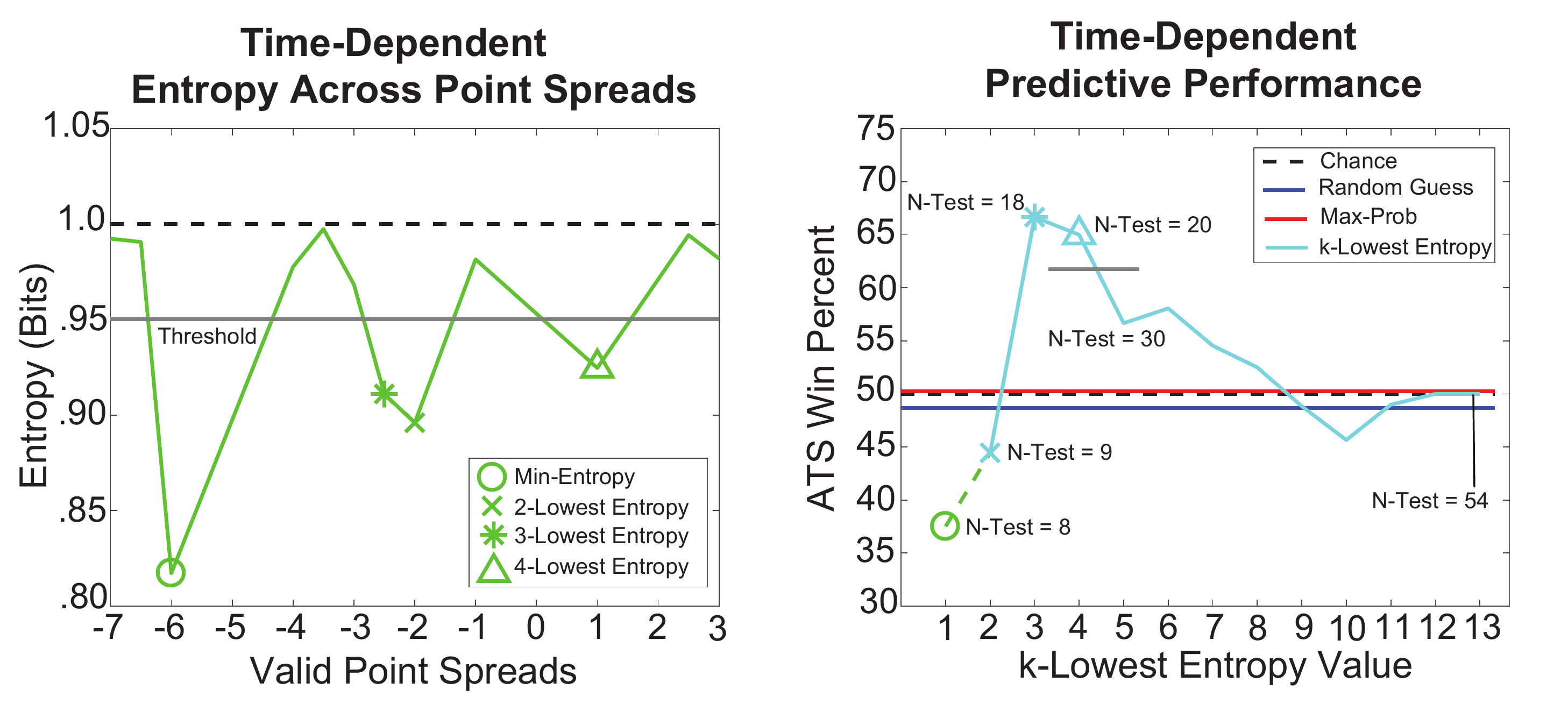}
\caption{Diagram depicting the entropy  across valid spreads \emph{(Left)} and the predictive performance of each strategy on test data under temporally-dependent conditions \emph{(Right)}.}
\label{fig:fig5}
\end{center}
\vskip -0.2in
\end{figure}

Figure \ref{fig:fig5} (Left) shows the entropy for each of the 12 valid spreads in the TD condition.  It shows there are 4 spreads that exhibit biases that exceed the entropy threshold used in this study (.95).  As a result, the k-Lowest Entropy model has a maximum $k=4$ spreads to wager across.

Surprisingly, the Min-Entropy model performed poorly (37.5\%) in percent ATS wins across the $n_t=8$ test samples that correspond to the Min-Entropy spread (Figure \ref{fig:fig5} (Right)). However, the k-Lowest Entropy method realized good predictive performance from $k=3$ (66.67\% ATS wins, $n_t=18$) and $k=4$ (65.00\% ATS wins, $n_t=20$) conditions.  The $k=2$ condition (44.44\% ATS wins, $n_t=9$) performed at similar levels to the Max-Probability (50.00\% ATS wins, $n_t=54$) and Random (48.68\%, $n_t=54$) models.

In order to visualize the impact on performance from increasing $k$ beyond the entropy threshold (.95), Figure \ref{fig:fig5} (Right) shows how ATS win percent \emph{decreases} for all spreads associated with $k>4$.  This suggests that performance gains were the result of exploiting spreads with significant decision bias.

\section{Discussion}

This effort was the first to investigate if oddsmaker decision biases can be exploited to improve the prediction of NFL game outcomes.  The study evaluated two models that do not exploit decision bias (Random-Guess and Max-Probability methods) against two approaches that exploit bias (Min-Entropy and k-Lowest Entropy) in their ability to predict the outcomes of NFL games.

The results show that under temporally-independent conditions, methods that exploit decision-bias predict NFL outcomes better than those that do not (Figure \ref{fig:fig4} (Right)).  However, when the same models were tested under conditions that account for the dates in which games were played (temporally-dependent), only the k-Lowest Entropy methods predicted NFL outcomes above chance (Figure \ref{fig:fig5} (Right)). For max-k conditions (TI: 2, TD:4), the k-Lowest Entropy method was able to achieve at least 65\% ATS win percent under both TI and TD conditions.

Taken together, it suggests that the k-Lowest Entropy method is a robust approach capable of improving predictive performance by exploiting oddsmaker decision biases.  This result is surprising, given the relative simplicity of the model (only using spread as a feature), compared to previous work in this area (\cite{Aoki:2017,David:11,Haghighat:13,Kiraly:2017,Knowlton:2017,Pelechrinis:2017}). This is likely due to the fact that in order to set spread-values, oddsmakers must account for many game and situational variables (Figure \ref{fig:fig1}).  Therefore, spread appears to be an extremely useful summary feature from which one can evaluate bias.

Another compelling issue to pursue is attempting to gain insight into the underlying reason oddsmakers exhibit biases.  As our analyses show (Figure \ref{fig:fig4}, (Left) and Figure \ref{fig:fig5}, (Left)), biases tend to be relatively stable for some spreads (e.g., -2.5).  This implies that either the algorithm or human is producing this spread for cases in which the visiting team wins frequently.  It's interesting to note looking at the marginal spread distribution ((Figure \ref{fig:fig2}), that the mean (-2.24) is near this maximum-bias location.  This may imply that this spread corresponds to the default \emph{home-field-advantage} produced when two teams are incorrectly estimated to be evenly matched.  Future work will explore this topic in greater detail.

Future work will also investigate more rigorous methods to set the entropy threshold used to identify biased spreads.  This value was set visually (at .95) in this study, so a quantitative threshold selection method is needed.  Moreover, efforts will be made to acquire greater amounts of data that can be used to further train and evaluate the models proposed in this study.

Overall, these results demonstrate how identifying and exploiting decision-biases can improve performance in competitive wagering situations.  Indeed, these results may extend to other competitive decision-making tasks, such as stock selection, where human and algorithmic decisions could be biased.  It will be interesting to see if this work can be generalized to other important domains.

\subsubsection*{Acknowledgments}

The author would like to thank his wife and son for their patience while this analysis was being performed.

\bibliography{schlicht_nfl}
\bibliographystyle{schlicht_nfl}

\newpage

\section*{Supplementary Figures}

\begin{suppfigure}[ht]
\vskip 0.2in
\begin{center}
\includegraphics[width=\columnwidth]{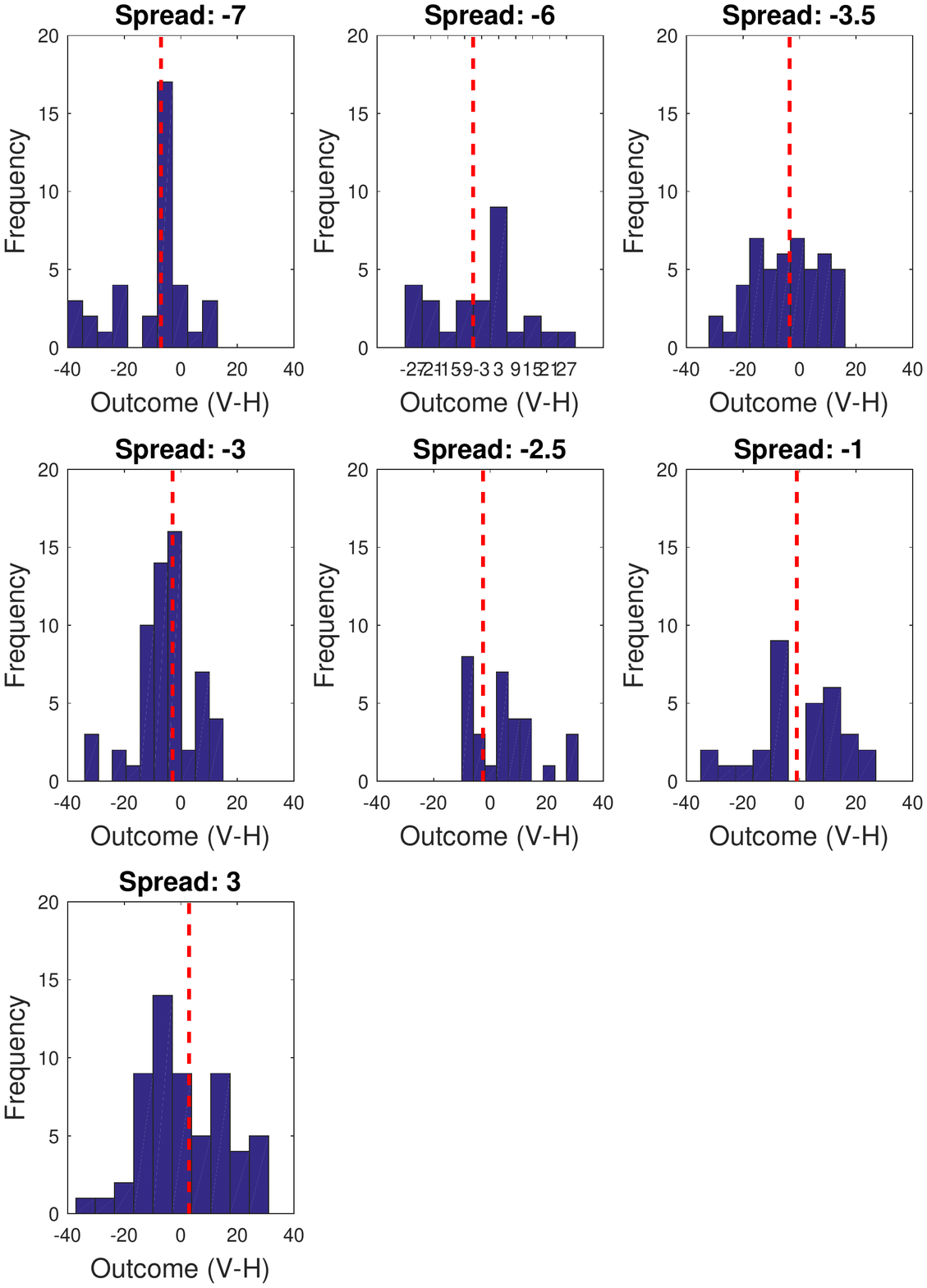}
\caption{Empirical outcome (visitor - home) histograms for each of the valid spread values used in the temporally-independent analysis}
\label{fig:figS1}
\end{center}
\vskip -0.2in
\end{suppfigure}

\newpage

\begin{suppfigure}[ht]
\vskip 0.2in
\begin{center}
\includegraphics[width=\columnwidth]{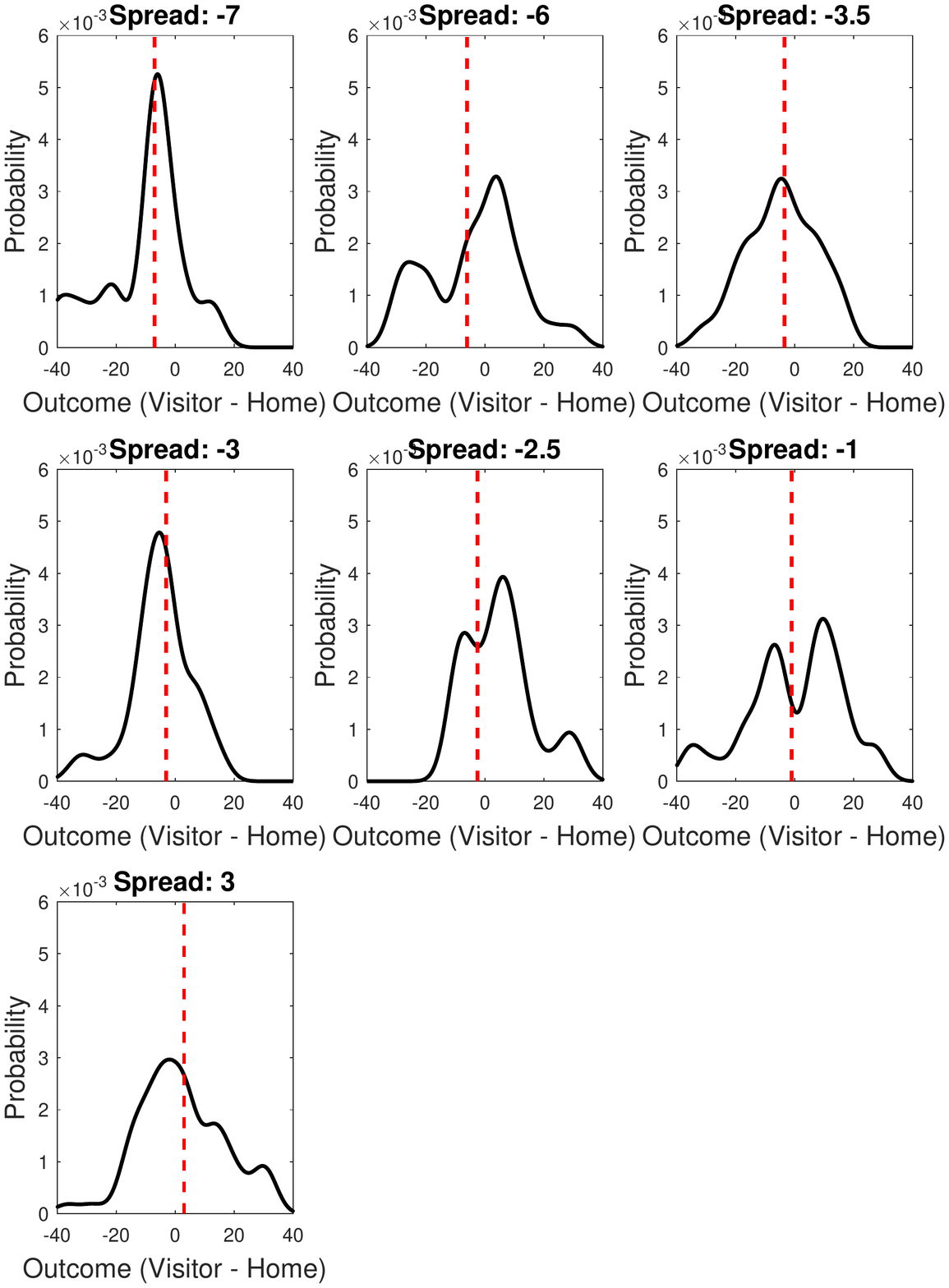}
\caption{Estimated pdf ($p(o \mid s)$) for each of the valid spread values used in the temporally-independent analysis}
\label{fig:figS2}
\end{center}
\vskip -0.2in
\end{suppfigure}

\newpage

\begin{table}[t]
  \caption{Supplementary Table}
  \label{ST1}
  \centering
  \begin{tabular}{lll}
    \toprule
    Model     & Percent ATS Win & N Test-Samples  \\
    \midrule
    Random       & 48.68\%      & 54     \\
    Max-Prob     &50.00\%       & 54      \\
    Min-Ent &37.50\%    & 8  \\
    2-Lowest Ent &44.44\%    & 9  \\
    3-Lowest Ent &66.67\%    & 18  \\
    4-Lowest Ent &65.00\%    & 20  \\
    5-Lowest Ent &56.67\%    & 30  \\
    6-Lowest Ent &58.06\%    & 31  \\
    7-Lowest Ent &52.50\%    & 33  \\
    8-Lowest Ent &48.84\%    & 40  \\
    9-Lowest Ent &45.65\%    & 46  \\
    10-Lowest Ent &48.98\%    & 49  \\
    11-Lowest Ent &50.00\%    & 52  \\
    12-Lowest Ent &50.00\%    & 54  \\
    \bottomrule
  \end{tabular}
\end{table}

\end{document}